\documentclass[aps,pra,superscriptaddress,reprint,twocolumn]{revtex4-1}
\usepackage{graphicx}
\begin{document}
\title{Role of generalized parity in the symmetry of fluorescence spectrum from two-level systems under periodic frequency modulation}
\date{\today}
\author{Yiying Yan}\email{yiyingyan@zust.edu.cn}
\affiliation{Department of Physics, School of Science, Zhejiang University of Science and Technology, Hangzhou 310023, China}
\author{Zhiguo L\"{u}}\email{zglv@sjtu.edu.cn}
\affiliation{Key Laboratory of Artificial Structures and Quantum Control (Ministry of Education), Department of Physics and Astronomy, Shanghai Jiao Tong University, Shanghai 200240, China}
\affiliation{Collaborative Innovation Center of Advanced Microstructures, Nanjing 210093, China}
\author{JunYan Luo}
\affiliation{Department of Physics, School of Science, Zhejiang University of Science and Technology, Hangzhou 310023, China}
\author{Hang Zheng}\email{hzheng@sjtu.edu.cn}
\affiliation{Key Laboratory of Artificial Structures and Quantum Control (Ministry of Education), Department of Physics and Astronomy, Shanghai Jiao Tong University, Shanghai 200240, China}
\affiliation{Collaborative Innovation Center of Advanced Microstructures, Nanjing 210093, China}
\begin{abstract}
We study the origin of the symmetry of the fluorescence spectrum from the two-level system subjected to a low-frequency periodic modulation and a near-resonant high-frequency monochromatic excitation by using the analytical and numerical methods based on the Floquet theory. We find that the fundamental origin of symmetry of the spectrum can be attributed to the presence of the generalized parity of the Floquet states, which depends on the driving parameters. The absence of the generalized parity can lead to the asymmetry of the spectrum. Based on the generalized parity, the conditions for the symmetry and asymmetry of the spectrum can be derived, which succeeds in predicting symmetry and asymmetry of the spectrum for the harmonic, biharmonic, and multiharmonic modulations. Moreover, we find that the secular approximation widely used in the analytical calculation may lead to artifact symmetry of the spectrum that vanishes when such approximation is avoided. The present study provides a significant perspective on the origin of the symmetry of the spectrum.
\end{abstract}
\maketitle

\section{Introduction}
Resonance fluorescence, arising from a quantum emitter driven by an external field and coupled to a radiative reservoir~\cite{PhysRev.188.1969,scully1997quantum,cohen1992atom}, is not only an important concept in quantum optics but also has potential application in quantum information technology, for instance, it plays an important role in realizing the single-photon source~\cite{he2013on-demand,PhysRevB.95.201410,PhysRevB.96.165306}. Particularly, the resonance fluorescence of two-level systems has attracted much interest and been studied in various aspects such as spectrum~\cite{PhysRevA.48.3092,Agarwal:91,PhysRevA.53.4275,PhysRevA.60.R4245,PhysRevB.89.155305,PhysRevA.90.023810,PhysRevLett.114.097402,PhysRevX.6.031004}, squeezing~\cite{PhysRevLett.55.2790,PhysRevLett.109.013601,PhysRevA.88.023837}, photon statistics~\cite{PhysRevLett.39.691,PhysRevA.41.4083,PhysRevB.78.153309,PhysRevA.90.063831}, photon antibunching~\cite{PhysRevA.38.559,PhysRevA.29.2004,PhysRevA.97.023832}, and so on. The line shape of the spectrum is found to depend strongly on the external field that interacts with the quantum emitters as well as the reservoirs to which the quantum emitters are coupled. As is well-known, for a sufficiently strong monochromatic field, the spectrum has a symmetric three-peak structure, known as the Mollow triplet~\cite{PhysRev.188.1969}. More recently, the bi- and multi-chromatically driven quantum systems are of interest~\cite{PhysRevA.95.013834,PhysRevA.96.063812,PhysRevA.97.033817,PhysRevA.98.043814}. In such systems, the spectrum turns out to have a complicated multipeak structure~\cite{PhysRevA.48.3092,Agarwal:91,PhysRevA.53.4275,PhysRevA.60.R4245,PhysRevB.89.155305,PhysRevA.90.023810,PhysRevLett.114.097402}, which can be either symmetric or asymmetric. In principle, the physical origin of the triplet and multipeak structures can be understood in terms of the transitions between the quantum dressed states~\cite{cohen1992atom} or in terms of the transitions between the semiclassical Floquet states~\cite{PhysRevA.55.3101,zheng2016}. The studies on the resonance fluorescence have enriched the physics concerning the light-matter interaction.

The origin of the symmetry of the spectrum has been found in the case of the monochromatic field. Specifically, it is the detailed balance condition that guarantees the symmetry of the Mollow triplet~\cite{cohen1992atom}. As is well-known, the breakdown of such a condition leads to the asymmetry of the spectrum, for instance, in the presence of a pure dephasing reservoir~\cite{PhysRevB.85.115309,PhysRevLett.110.217401} or the counter-rotating terms of the external field under certain conditions~\cite{Browne,PhysRevA.88.053821,zheng2016}. The dephasing-induced asymmetric Mollow triplet has been experimentally observed in the quantum dots (the pure dephasing arises because of the interaction between the quantum dot and its solid-state environment)~\cite{PhysRevLett.106.247402,Ulhaq:13}. For the bi- and multi-chromatic fields, the origin of the symmetry of the spectrum is rarely discussed, owing to the fact that the physically transparent spectrum is hardly analytically derived, and has not been comprehensively understood.

Recent studies show that the fluorescence spectrum from a driven two-level system with a modulated transition frequency is symmetrically multipeaked for the vanishing detuning while asymmetrically multipeaked for the finite detuning~\cite{PhysRevA.93.033812,PhysRevA.95.013834,PhysRevA.96.063812,PhysRevA.97.033817}. Such an exotic bichromatically driven two-level system with coexistence of the longitudinal and transversal coupling between the system and the applied fields has been experimentally studied in the superconducting qubits~\cite{li2013motional,PhysRevB.96.174518}, single molecule~\cite{PhysRevLett.81.2679}, and nitrogen-vacancy spin qubits~\cite{PhysRevLett.112.010502}. The quantum systems under frequency modulation are also of interest in theoretical studies~\cite{PhysRevLett.102.023601,PhysRevA.90.043838,PhysRevA.91.053820,PhysRevB.87.134505,PhysRevA.92.013846}, the intriguing phenomena of which were reviewed recently~\cite{Silveri2017Quantum}. It is worthwhile to note that the bichromatically driven two-level system with frequency modulation differs from those considered in Refs.~\cite{Agarwal:91,PhysRevA.48.3092}, where the two-level systems are transversely driven by a bichromatic field. In such a case, the symmetry of the fluorescence spectrum is found to depend on the average detuning if the strengths of the two components of the bichromatic field are the same; the pronounced asymmetry of the spectrum is revealed when the average detuning is finite and/or the strengths of the two components of the field are unequal~\cite{Agarwal:91,PhysRevA.48.3092}.
For a bichromatically amplitude-modulated field, the spectrum is also found to be symmetric and asymmetric for the vanishing and finite detuning, respectively~\cite{PhysRevA.40.3164}. So far the fundamental origin of such a detuning-dependent symmetry remains obscure.

In this work, we use both analytical and numerical methods based on the Floquet theory to study the fundamental origin of the symmetry of the fluorescence spectrum from the two-level system under a low-frequency periodic modulation and a near-resonant monochromatic excitation. We address the symmetry and asymmetry of the spectrum by considering the generalized parity of Floquet states rather than the behaviors of the bare-state or dressed-state populations as considered in Refs.~\cite{PhysRevB.88.125306,PhysRevA.92.013846,PhysRevA.96.063812}. The generalized parity is found to guarantee the symmetry of the spectrum while the breaking of such a parity can yield pronouncedly asymmetric spectrum even in the vanishing detuning case. Based on the generalized parity, the conditions for the symmetric and asymmetric spectra are derived, which are not given in the previous works and cannot be derived from the behaviors of the bare or dressed state population. The generalized-parity-induced symmetry of the spectrum is verified and illustrated in the context of the biharmonic modulation by the comparison between the analytical and numerical results. The analytical results are found to be in agreement with the numerically exact results in the regimes where the perturbation theory and secular approximation can be justified. In addition, we find that the spectrum with the secular approximation may have artifact symmetry under certain conditions, i.e., the spectrum with secular approximation is symmetric while the numerically exact calculation shows asymmetric spectra because of the broken parity. The present finding simply interprets the detuning-dependent symmetry in the harmonic modulation case and can also be extended to analyze the symmetry and asymmetry of the spectrum in the multiharmonic modulation cases. Our results suggest that it is feasible to control the symmetry and asymmetry of the spectrum via engineering the generalized parity of the Floquet states.

The rest of the paper is organized as follows. In Sec.~\ref{sec:fsgp}, we first discuss the generalized-parity-induced symmetry of the fluorescence spectrum without the secular approximation and further elucidate the symmetry of the spectrum with a physically transparent formal spectrum with the secular approximation. In Sec.~\ref{sec:verfication}, we analytically and numerically calculate the fluorescence spectrum in the context of the biharmonic modulation to verify the symmetry and asymmetry of the spectrum predicted based on the generalized parity. In the last section, the conclusions are given.

\section{Fluorescence spectrum and generalized parity}\label{sec:fsgp}

We consider that the transition frequency of the two-level system
is modulated periodically via a low-frequency external field $f(t)$
and the two-level system is also excited by a near-resonant monochromatic
field, which is described by the following Hamiltonian ($\hbar=1$)
\begin{equation}
H(t)=\frac{1}{2}[\omega_{0}+f(t)]\sigma_{z}+\frac{\Omega_{x}}{2}(\sigma_+e^{-i\omega_x t}+\sigma_-e^{i\omega_x t}),
\end{equation}
where $\sigma_{z(x,y)}$ is the usual Pauli matrix, $\omega_{0}+f(t)$ is
the modulated transition frequency, $\sigma_{\pm}=(\sigma_{x}\pm i\sigma_{y})/2$
are the raising and lowering operators, and $\Omega_{x}$ ($\omega_{x}$)
is the strength (frequency) of the monochromatic driving. Here we choose
$f(t)=f(t+T)$ with $T$ being the fundamental period of the modulation and
much greater than $2\pi/\omega_{x}$. This is a generalized model as compared with the previous one considered in Refs.~\cite{PhysRevA.93.033812,PhysRevA.95.013834,PhysRevA.96.063812}.

To study the emission processes, we need to take account of the spontaneous decay. Thus, the time evolution of the driven two-level system under study
is modeled by the Lindblad master equation.
In the frame rotating at the frequency $\omega_x$, the Lindblad master equation takes the form
\begin{equation}
\frac{d}{dt}\tilde{\rho}(t)={\cal L}(t)\tilde{\rho}(t),\label{rotateme}
\end{equation}
where $\tilde{\rho}(t)$ is the reduced density matrix in the
rotating frame and the superoperator ${\cal L}(t)$ is given by ${\cal L}(t)\tilde{\rho}(t)=-i[\tilde{H}(t),\tilde{\rho}(t)]-\kappa/2[\{\sigma_{+}\sigma_{-},\tilde{\rho}(t)\}-2\sigma_{-}\tilde{\rho}(t)\sigma_{+}]$ with $\kappa$ being the radiative decay rate. $\tilde{H}(t)$ is the effective Hamiltonian and reads
\begin{equation}
\tilde{H}(t)=\frac{\Omega_{x}}{2}\sigma_{x}+\frac{1}{2}[\delta+f(t)]\sigma_{z},\label{eq:HRWA}
\end{equation}
with $\delta=\omega_{0}-\omega_{x}$ being the detuning between the
bare transition frequency and monochromatic excitation frequency.
This master equation is actually a set of first-order
differential equations with periodic coefficients. It can be
directly solved by the so-called Floquet-Liouville (FL) approach with a desire accuracy~\cite{PhysRevA.33.1798,PhysRevA.93.033812}.
Although such a Floquet-theory-based numerical method is simple and efficient, it
is not physically transparent to analyze the role of generalized parity
of Floquet states in the symmetry of the fluorescence spectrum.
We use an alternative method which is developed in our previous works~\cite{zheng2016,PhysRevA.97.033817} to solve the master equation and calculate
the fluorescence spectrum.
We first calculate the Floquet states for $\tilde{H}(t)$ and use
them as the bases to reformulate Eq.~(\ref{rotateme}) and derive its
analytical formal solutions with the aid of the secular approximation in the Floquet
picture.

\subsection{The symmetry of fluorescence spectrum without secular approximation}

The steady-state fluorescence spectrum is given by the Fourier transform
of the time-averaged first-order correlation function~\cite{PhysRev.188.1969,PhysRevA.33.1798}
\begin{equation}
S(\Delta)\propto{\rm Re}\frac{1}{T}\int_{0}^{\infty}\int_{0}^{T}\lim_{t^{\prime}\rightarrow\infty}\left\langle \tilde{\sigma}_{+}(t^{\prime}+\tau)\tilde{\sigma}_{-}(t^{\prime})\right\rangle e^{-i\Delta\tau}dt^{\prime}d\tau,
\end{equation}
where $\Delta=\omega-\omega_{x}$ and $\left\langle \tilde{\sigma}_{+}(t^{\prime}+\tau)\tilde{\sigma}_{-}(t^{\prime})\right\rangle $
is the first-order correlation function and the tilde indicates that
it is evaluated in the rotating frame. In general, it is difficult to derive an exact analytical spectrum. Nevertheless, we find that it is possible to show that the spectrum is exactly symmetric about $\Delta=0$ when $\delta+f(t)=-[\delta+f(t+T/2)]$ by realizing the fact that the driven two-level system possesses a generalized parity symmetry, i.e.,
\begin{equation}
  \sigma_x\tilde{H}(t+T/2)\sigma_x=\tilde{H}(t). \label{transHt}
\end{equation}
Here, the generalized parity transformation consists of an exchange between the up and down states
of two-level system ($\sigma_z\rightarrow-\sigma_z$) and a time shift of half period of the modulation ($t\rightarrow t+T/2$).

We state briefly how the generalized parity guarantees the symmetry of the spectrum. Owing to Eq.~(\ref{transHt}), we can construct a generalized parity transformation in the Liouville space, the details of which can be found in Appendix~\ref{AppC}. When $\delta+f(t)=-[\delta+f(t+T/2)]$, the superoperator ${\cal L}(t)$ is similarly found to be invariant under the generalized parity transformation. Based on this property, it can be derived from the master equation~(\ref{rotateme}) without the secular approximation that in the steady-state limit, the time-averaged first-order correlation function is a real-valued function in the rotating frame. As a result, the fluorescence spectrum is symmetric about $\Delta=0$. This finding shows that the symmetry of the spectrum occurs when $\delta+f(t)=-[\delta+f(t+T/2)]$ and results from the generalized parity. We will numerically verify the generalized-parity-induced symmetry in Sec.~\ref{sec:verfication}.

\subsection{The symmetry of fluorescence spectrum with secular approximation}

To further elucidate the role of the generalized parity in determining the symmetry of the spectrum, we calculate the spectrum in the Floquet picture which allows us to derive a physically transparent formal spectrum with the aid of the secular approximation.

According
to the Floquet theory~\cite{PhysRev.138.B979,PhysRevA.7.2203}, the time-dependent Schr\"{o}dinger equation
governed by $\tilde{H}(t)$ possesses a set of formal solutions $|\tilde{\psi}_{\alpha}(t)\rangle=|\tilde{u}_{\alpha}(t)\rangle e^{-i\tilde{\varepsilon}_{\alpha}t}$,
where $|\tilde{u}_{\alpha}(t)\rangle=|\tilde{u}_{\alpha}(t+T)\rangle$
is Floquet state and $\tilde{\varepsilon}_{\alpha}$ is the corresponding real-valued
quasienergy. The index $\alpha$ labels independent Floquet states. Substituting the formal solution into the Schr\"{o}dinger
equation, one readily finds that
\begin{equation}
[\tilde{H}(t)-i\partial_{t}]|\tilde{u}_{\alpha}(t)\rangle=\tilde{\varepsilon}_{\alpha}|\tilde{u}_{\alpha}(t)\rangle.\label{eigEQ}
\end{equation}
On solving this equation, one obtains the Floquet states and quasienergies of the driven two-level system.

We use $|\tilde{u}_{\alpha}(t)\rangle$ ($\alpha=\pm$) as the basis to reformulate
the master equation~(\ref{rotateme}) and invoke the secular approximation~\cite{zheng2016,PhysRevA.97.033817},
yielding
\begin{eqnarray}
\frac{d}{dt}\tilde{\rho}_{++}(t) & = & -\Gamma_{{\rm rel}}\tilde{\rho}_{++}(t)+\Gamma_{{\rm s}},\\
\frac{d}{dt}\tilde{\rho}_{+-}(t) & = & -(i\Delta_{+-}+\Gamma_{{\rm deph}})\tilde{\rho}_{+-}(t),
\end{eqnarray}
where $\tilde{\rho}_{\alpha\beta}(t)=\langle\tilde{u}_{\alpha}(t)|\tilde{\rho}(t)|\tilde{u}_{\beta}(t)\rangle$
is the element of density operator, $\Delta_{+-}=\tilde{\varepsilon}_{+}-\tilde{\varepsilon}_{-}$
is the difference of two quasienergies, and $\Gamma_{{\rm s}}=\kappa\sum_{l}|x_{-+,l}^{(+)}|^{2}$, where $x^{(+)}_{\alpha\beta,l}$ is a time-averaged transition matrix element defined as follows:
\begin{equation}
  x^{(\pm)}_{\alpha\beta,l}=\frac{1}{T}\int^{T}_{0}\langle\tilde{u}_\alpha(t)|\sigma_\pm|\tilde{u}_{\beta}(t)\rangle e^{-i2\pi lt/T}dt. \label{eq:xabl}
\end{equation}
The relaxation rate $\Gamma_{{\rm rel}}$ and dephasing rate $\Gamma_{{\rm deph}}$ are given by
\begin{eqnarray}
\Gamma_{{\rm rel}} & = & \kappa\sum_{l}(|x_{+-,l}^{(+)}|^{2}+|x_{-+,l}^{(+)}|^{2}),\\
\Gamma_{{\rm deph}} & = & \frac{\kappa}{2}\sum_{l}(|x_{+-,l}^{(+)}|^{2}+|x_{-+,l}^{(+)}|^{2}+4|x_{++,l}^{(+)}|^{2}).
\end{eqnarray}
The analytical formal solutions in the Floquet picture
can be easily found as follows:
\begin{eqnarray}
\tilde{\rho}_{++}(t) & = & \tilde{\rho}_{++}(0)e^{-\Gamma_{{\rm rel}}t}+\tilde{\rho}_{++}^{{\rm ss}}(1-e^{-\Gamma_{{\rm rel}}t}),\\
\tilde{\rho}_{+-}(t) & = & \tilde{\rho}_{+-}(0)e^{-(\Gamma_{{\rm deph}}+i\Delta_{+-})t},
\end{eqnarray}
where
\begin{equation}
\tilde{\rho}_{++}^{{\rm ss}}=\frac{\Gamma_{{\rm s}}}{\Gamma_{{\rm rel}}}=\frac{\sum_{l}|x_{-+,l}^{(+)}|^{2}}{\sum_{l}(|x_{+-,l}^{(+)}|^{2}+|x_{-+,l}^{(+)}|^{2})}\label{eq:rhopp}
\end{equation}
 is the steady-state population of the Floquet state. These solutions
together with the quantum regression theory enable us to
derive a physically transparent spectrum function~\cite{zheng2016,PhysRevA.97.033817}
\begin{eqnarray}
S(\Delta) & \propto & \sum_{l}\bigg\{\pi|x_{++,l}^{(+)}|^{2}(\tilde{\rho}_{++}^{{\rm ss}}-\tilde{\rho}_{--}^{{\rm ss}})^{2}\delta(\Delta-l\omega_{z})\nonumber \\
 &  & +4|x_{++,l}^{(+)}|^{2}\tilde{\rho}_{++}^{{\rm ss}}\tilde{\rho}_{--}^{{\rm ss}}\frac{\Gamma_{{\rm rel}}}{\Gamma_{{\rm rel}}^{2}+(\Delta-l\omega_{z})^{2}}\nonumber \\
 &  & +|x_{+-,l}^{(+)}|^{2}\tilde{\rho}_{++}^{{\rm ss}}\frac{\Gamma_{{\rm deph}}}{\Gamma_{{\rm deph}}^{2}+(\Delta-l\omega_{z}-\Delta_{+-})^{2}}\nonumber \\
 &  & +|x_{-+,l}^{(+)}|^{2}\tilde{\rho}_{--}^{{\rm ss}}\frac{\Gamma_{{\rm deph}}}{\Gamma_{{\rm deph}}^{2}+(\Delta-l\omega_{z}+\Delta_{+-})^{2}}\bigg\},\nonumber\\\label{eq:sfunsa}
\end{eqnarray}
It is evident that the accuracy of Eq.~(\ref{eq:sfunsa}) is limited by the secular approximation when the transition matrix elements $x^{(+)}_{\alpha\beta,l}$ and quasienergies are exactly calculated. As is well-known, the secular approximation can be justified under the strong driving condition, i.e., $\Delta_{+-}\gg\kappa$. In general, we can calculate the quasienergies and transition matrix elements based on both analytical and numerical diagonalization (ND) of the Floquet Hamiltonian $\tilde{H}(t)-i\partial_t$ in the Sambe space~\cite{PhysRev.138.B979,PhysRevA.7.2203} , yielding the analytical and semianalytical spectra, respectively.

Next, we discuss the parity phenomenon of the Floquet states resulting from Eq.~(\ref{transHt}). We consider the behavior of the Floquet
states under the generalized parity transformation ${\cal P}_G$, which is defined as
\begin{equation}
{\cal P}_{G}|\tilde{u}_{\alpha}(t)\rangle:=\sigma_{x}|\tilde{u}_{\alpha}(t+T/2)\rangle.\label{eq:hpt}
\end{equation}
By differentiating $\sigma_{x}|\tilde{u}_{\alpha}(t+T/2)\rangle$
with respect to $t$, we readily obtain
\begin{widetext}
\begin{equation}
\left[\sigma_{x}\tilde{H}\left(t+T/2\right)\sigma_{x}-i\partial_{t}\right]\sigma_{x}\left|\tilde{u}_{\alpha}\left(t+T/2\right)\right\rangle =\tilde{\varepsilon}_{\alpha}\sigma_{x}\left|\tilde{u}_{\alpha}\left(t+T/2\right)\right\rangle .
\end{equation}
\end{widetext}
When $\delta+f(t)=-[\delta+f(t+T/2)]$, $\sigma_{x}|\tilde{u}_{\alpha}(t+T/2)\rangle$ satisfies the same differential equation as $|\tilde{u}_{\alpha}(t)\rangle$ because of Eq.~(\ref{transHt}).
Recalling the uniqueness of solutions of the differential equations, in such cases
we must have
\begin{equation}
\sigma_{x}\left|\tilde{u}_{\alpha}\left(t+T/2\right)\right\rangle =\lambda_{\alpha}|\tilde{u}_{\alpha}(t)\rangle,\label{eq:parity}
\end{equation}
where $\lambda_{\alpha}$ is a constant. Furthermore, we have $\lambda_{\alpha}=\pm1$
because of ${\cal P}_{G}^{2}|\tilde{u}_{\alpha}(t)\rangle=\lambda_{\alpha}^{2}|\tilde{u}_{\alpha}(t)\rangle=|\tilde{u}_{\alpha}(t)\rangle.$ Specifically, when $\delta+f(t)=-[\delta+f(t+T/2)]$, the Floquet states may be even or odd functions under the generalized parity transformation, which is referred to as the generalized parity of the Floquet states. The generalized parity has been previously investigated in other phenomena such as the coherent destruction of
tunneling~\cite{PhysRevLett.67.516} and the laser-induced electronic transport~\cite{doi:10.1063}.

Clearly, if $\delta+f(t)\neq-\left[\delta+f\left(t+T/2\right)\right]$,
Eq.~(\ref{eq:parity}) cannot hold as $\sigma_{x}\tilde{H}\left(t+T/2\right)\sigma_{x}\neq\tilde{H}(t)$,
i.e., the effective Hamiltonian is no longer invariant under the generalized parity
transformation. Consequently, the Floquet states also do not have the generalized parity.

\begin{figure*}
  \includegraphics[width=2\columnwidth]{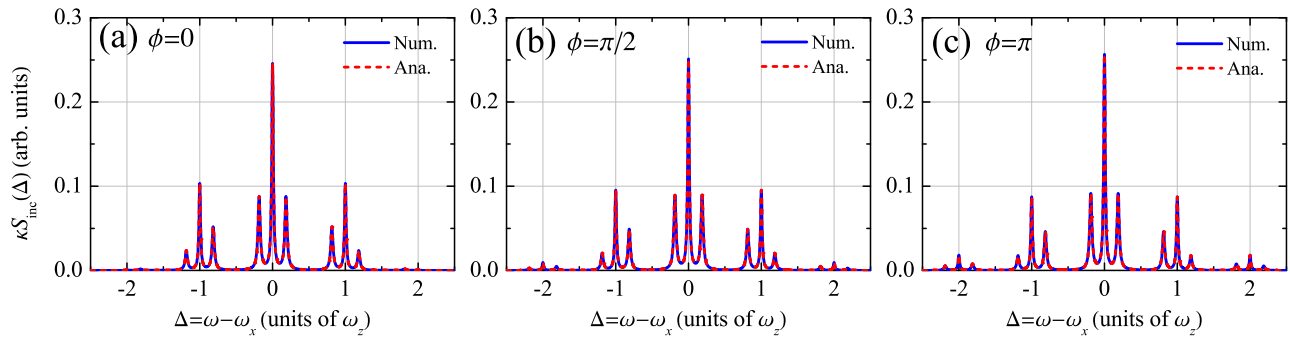}
  \caption{The incoherent components of the fluorescence spectrum for $p=3$, $\Omega_x=10\kappa$, $\delta=0$, $\Omega_z=\omega_z=40\kappa$, $r=1$, and various phase. ``Ana.'' and ``Num.'' denote the analytical and the FL numerical results, respectively.}\label{fig1}
\end{figure*}

We show that the symmetry of the spectrum may be a consequence of the generalized parity of the Floquet states. By using Eq.~(\ref{eq:parity}) and $x_{\alpha\beta,l}^{(+)}=\left[x_{\beta\alpha,-l}^{(-)}\right]^{\ast}$,
it is straightforward to show the following identity for arbitrary integer $l$ from the definition~(\ref{eq:xabl}) of the transition matrix element:
\begin{equation}
x_{\alpha\beta,l}^{(+)}=(-1)^{l}\lambda_{\alpha}\lambda_{\beta}\left[x_{\beta\alpha,-l}^{(+)}\right]^{\ast},\label{eq:identity}
\end{equation}
provided $\delta+f(t)=-[\delta+f(t+T/2)]$. It follows that
\begin{equation}
  |x_{\alpha\beta,l}^{(+)}|=|x_{\beta\alpha,-l}^{(+)}|\label{eq:identity2}
\end{equation}
also holds for any integer $l$. We emphasize that the relation~(\ref{eq:identity2}) can be deduced from relation~(\ref{eq:identity}), however, the relation~(\ref{eq:identity}) cannot be derived from relation~(\ref{eq:identity2}). With the relation~(\ref{eq:identity2}), it is straightforward to show that the spectrum~(\ref{eq:sfunsa}) is symmetric about $\Delta=0$~\cite{PhysRevA.97.033817}. Specifically, since
$|x_{++,l}^{(+)}|=|x_{++,-l}^{(+)}|$,
the emission lines at $\Delta=\pm l\omega_{z}$
(the positions are symmetric about $\Delta=0$) have the
equal weights. Moreover, since $|x_{+-,l}^{(+)}|=|x_{-+,-l}^{(+)}|$,
we also have $\tilde{\rho}_{++}^{{\rm ss}}=\tilde{\rho}_{--}^{{\rm ss}}$
according to Eq.~(\ref{eq:rhopp}), leading to
$|x_{+-,l}^{(+)}|^{2}\tilde{\rho}_{++}^{{\rm ss}}=|x_{-+,-l}^{(+)}|^{2}\tilde{\rho}_{--}^{{\rm ss}}$.
That is to say, the emission lines at $\Delta=\pm(l\omega_{z}+\Delta_{+-})$
(the positions are symmetric about $\Delta=0$) have the
same weights. It turns out that the symmetry of the spectrum fundamentally originates from the generalized parity of the Floquet states when $\delta+f(t)=-[\delta+f(t+T/2)]$. Conversely, one may expect that the symmetry of the spectrum may break when such a parity is absent. However, it is a formidable task to analytically prove that the spectrum is asymmetric in the absence of the generalized parity.

Let us discuss what happens to the formal spectrum if $\delta+f(t)\neq-[\delta+f(t+T/2)]$. Under such a condition, the generalized parity is absent, and thus we cannot have the relation~(\ref{eq:identity}). In principle, the absence of the generalized parity will result in two possible situations. One is that the spectrum becomes asymmetric about $\Delta=0$ because the relation $|x^{(+)}_{\alpha\beta,l}|\neq|x^{(+)}_{\beta\alpha,-l}|$ can be derived at least for a certain $l$. The other is that the spectrum is symmetric because the equality $|x^{(+)}_{\alpha\beta,l}|=|x^{(+)}_{\beta\alpha,-l}|$ still holds for any $l$, originating from other kinds of identities between the transition matrix elements rather than the generalized-parity-induced identity~(\ref{eq:identity}). Apparently the first situation is more trivial than the second one. Most importantly, the present analysis suggests that
the formal spectrum may be symmetric even without the generalized parity. Consequently, we cannot conclude from the formal spectrum~(\ref{eq:sfunsa}) that the symmetry of the spectrum breaks as long as the generalized parity is absent.

To end this section, we give some remarks on the above findings based on the formal spectrum. First, we find that the symmetry of the spectrum may result from the generalized parity and requires $\delta+f(t)=-[\delta+f(t+T/2)]$. This is consistent with the analysis above without the secular approximation. Moreover, the generalized parity is found to be an important underlying cause of the relation~(\ref{eq:identity2}), which was numerically found in harmonic modulation case~\cite{PhysRevA.97.033817}. It turns out here that the relation~(\ref{eq:identity2}) can be established due to the generalized parity in the bi- and multi-harmonic cases. Second, without the generalized parity, namely, when $\delta+f(t)\neq-[\delta+f(t+T/2)]$, the formal spectrum can be either trivially asymmetric or nontrivially symmetric. The symmetry requires the relation~(\ref{eq:identity2}) in the absence of the generalized parity, namely, Eq.~(\ref{eq:identity}). Third, the formal spectrum is derived with the secular approximation and thus the present analysis needs further verification. In what follows we consider a concrete biharmonic modulation to verify whether the generalized parity guarantees the symmetry of the spectrum when the secular approximation is not invoked and we also check whether the relation~(\ref{eq:identity2}) can be established without the generalized parity and whether such relations lead to the symmetry of the spectrum without the secular approximation.

\begin{figure*}
  \includegraphics[width=2\columnwidth]{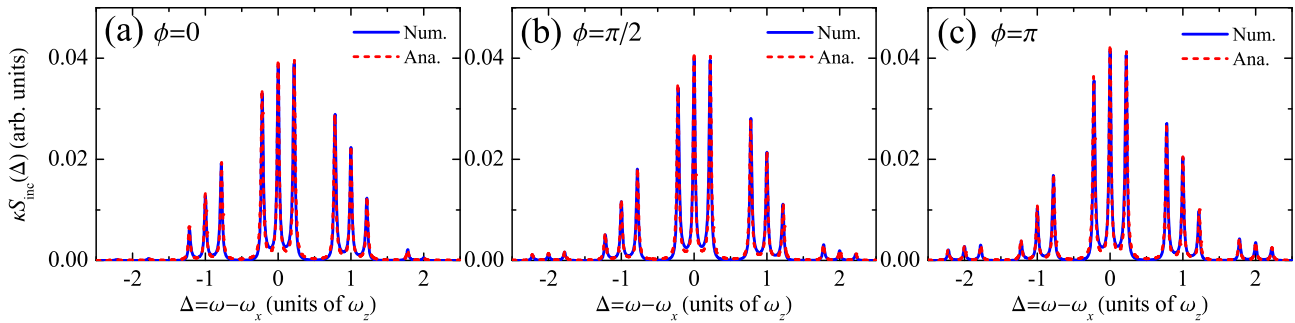}
  \caption{The incoherent components of the fluorescence spectrum for $p=3$, $\delta=5\kappa$, $\Omega_x=10\kappa$, $\Omega_z=\omega_z=40\kappa$, $r=1$, and various phase.}\label{fig2}
\end{figure*}

\section{Verification of symmetry and asymmetry of the spectrum}\label{sec:verfication}

To calculate fluorescence spectrum, without loss of generality, we mainly consider the biharmonic modulation in this work, namely,
the modulation consists
of two harmonics
\begin{equation}
f(t)=\Omega_{z}[\cos(\omega_{z}t)+r\cos(p\omega_{z}t+\phi)],\label{eq:bih}
\end{equation}
where $\Omega_{z}$ and $\omega_{z}=2\pi/T$ are the amplitude and fundamental frequency of
the modulation, respectively, $p$ is a positive integer, $r$ is the ratio of the amplitude of the second
harmonic to that of the first one, and $\phi$ is a relative phase. Since $\frac{1}{T}\int^{T}_{0}f(t)dt=0$, the condition for the presence of the generalized parity $\delta+f(t)=-[\delta+f(t+T/2)]$ is equivalent to $\delta=0$ and $f(t)=-f(t+T/2)$. The condition for the absence of the generalized parity $\delta+f(t)\neq-[\delta+f(t+T/2)]$ is simply divided into three cases:
\begin{equation}
\left\{ \begin{array}{c}
\delta\neq0\,{\rm and}\,f(t)=-f(t+T/2);\\
\delta=0\,{\rm and}\,f(t)\neq-f(t+T/2);\\
\delta\neq0\,{\rm and}\,f(t)\neq-f(t+T/2).
\end{array}\right.
\end{equation}
It is noted that for the biharmonic modulation~(\ref{eq:bih}), both $f(t)=-f(t+T/2)$ and $f(t)\neq-f(t+T/2)$ can be realized by setting $p$ odd and even numbers, respectively.
To verify above analysis, we calculate the numerically exact fluorescence spectrum from master equation~(\ref{rotateme}) with the FL formalism~\cite{PhysRevA.33.1798,PhysRevA.93.033812}, which is compared with the analytical and semianalytical results from Eq.~(\ref{eq:sfunsa}). The analytical and semianalytical results are obtained by using the transition matrix elements and quasienergies calculated with the Van Vleck perturbation theory and the ND of the Floquet Hamiltonian, respectively. The detailed analytical calculation is presented in Appendix~\ref{AppA}. In addition, we just focus on the incoherent components of the fluorescence spectrum, which is of interest in the experiments. In principle, similar analysis is applicable to the coherent components. In this work, we mainly consider the parameters regime $\omega_z\sim\Omega_z\gg\Omega_x\gg\kappa$, in which case both the Van Vleck perturbation theory (up to second order in $\Omega_x$) and secular approximation can be justified. Importantly, this regime is experimentally accessible in the artificial atoms, e.g., the transmon qubit~\cite{li2013motional}. We should emphasize that if the perturbation theory is inapplicable, we can obtain the transition matrix elements and quasienergies by the ND of the Floquet Hamiltonian.

We first verify whether the generalized parity guarantees the symmetry of the spectrum. In Fig.~\ref{fig1}, we display the incoherent
component of fluorescence spectra obtained by the FL numerical method (solid
line) and analytical result (dashed line) for $p=3$, $\delta=0$,
and various values of $\phi$. Apparently the spectra
are symmetric as expected. The analytical results are in agreement with
the FL results. These results also show that the spectrum
depends weakly on the relative phase $\phi$. In addition, it is straightforward to verify
that for other driving parameters, the spectrum is symmetric as well when $p$ is an odd number and
$\delta=0$. In Appendix~\ref{AppB}, we show that when $\delta=0$ and $p$ is odd, the transition matrix elements indeed satisfy Eq.~(\ref{eq:identity}), which guarantees the symmetry of the spectrum. The present results suggest that the symmetry of the spectrum appears as
long as $\delta=0$ and $f(t)=-f(t+T/2)$ and fundamentally originates from the
generalized parity of the Floquet states in such a situation.

\begin{figure*}
  \includegraphics[width=2\columnwidth]{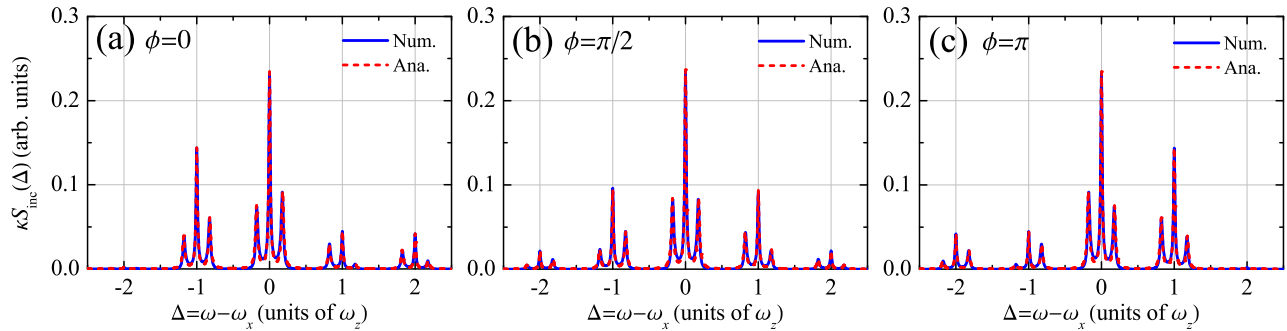}
  \caption{The incoherent components of the fluorescence spectrum for $p=2$, $\delta=0$, $\Omega_x=10\kappa$, $\Omega_z=\omega_z=40\kappa$, $r=1$, and various phases.}\label{fig3}
\end{figure*}

We move to examine whether the symmetry of the spectrum breaks when the generalized parity is absent, namely, under the conditions $\delta+f(t)\neq-[\delta+f(t+T/2)]$. We calculate the spectra with the parameters being
the same as in Fig.~\ref{fig1} except for the detuning $\delta=5\kappa$, corresponding to the case of $\delta\neq0$ and $f(t)=-f(T+T/2)$. In
Fig.~\ref{fig2}, the analytical and FL numerical spectra agree with each other and are found to be asymmetric for the finite detuning, indicating
that in spite of $f(t)=-f(t+T/2)$, the asymmetry of spectrum appears
when $\delta\neq0$.

Let us consider the case of $\delta=0$ and $f(t)\neq-f(t+T/2)$ by setting $p$ being even. We calculate the spectrum
for $p=2$ and the
other parameters being the same as in Fig.~\ref{fig1}. Figure~\ref{fig3} displays that
the analytical and numerical spectra are pronouncedly asymmetric even though $\delta=0$ except for $\phi=\pi/2$ in which case the
analytical spectrum is found to be
strictly symmetric (see discussion below) while the numerical spectrum is slightly asymmetric
[in particular, the intensities
of emission lines at $\Delta=\pm\omega_{z}$ are unequal as shown in Fig.~\ref{fig6}(a)]. These results
confirm that the formal spectrum~(\ref{eq:sfunsa}) may be symmetric without the generalized parity of the Floquet states. However, the numerically exact spectrum is asymmetric in the absence of the generalized parity. This shows that the generalized parity plays an important role in determining the symmetry of the exact spectrum. We will further analyze such discrepancy between the analytical and numerical results later. In addition, we find that in contrast with $p=3$, the spectrum is found to depend strongly on
relative phase $\phi$ when $p=2$.

\begin{figure*}
  \includegraphics[width=2\columnwidth]{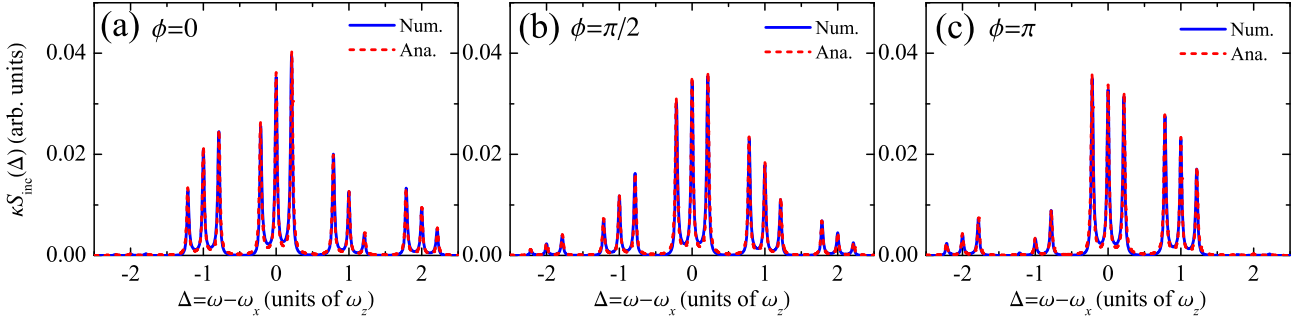}
  \caption{The incoherent components of fluorescence spectrum for $p=2$, $\delta=5\kappa$, $\Omega_x=10\kappa$, $\Omega_z=\omega_z=40\kappa$, $r=1$, and various phase.}\label{fig4}
\end{figure*}

Finally we calculate the spectra for $\delta\neq0$ and $f(t)\neq-f(t+T/2)$. Figure~\ref{fig4} shows
the spectra obtained for the detuning $\delta=5\kappa$
and the other parameters being the same as in Fig.~\ref{fig3}. The spectra are still asymmetric. In general, it is straightforward
to verify the asymmetry of the spectrum under the condition that $\delta+f(t)\neq-[\delta+f(t+T/2)]$. All in all, it turns out that the symmetry of the spectrum breaks in the absence of the generalized parity. Conversely, we can say that the symmetry of the
spectrum can be fully attributed to the presence of the
generalized parity. In contrast to the previous studies, we ascribe the asymmetry to the breaking of the generalized parity rather than the unequal populations of dressed states~\cite{PhysRevA.96.063812} or the breakdown of relation~(\ref{eq:identity2})~\cite{PhysRevA.97.033817}.

\begin{figure}
  \includegraphics[width=1\columnwidth]{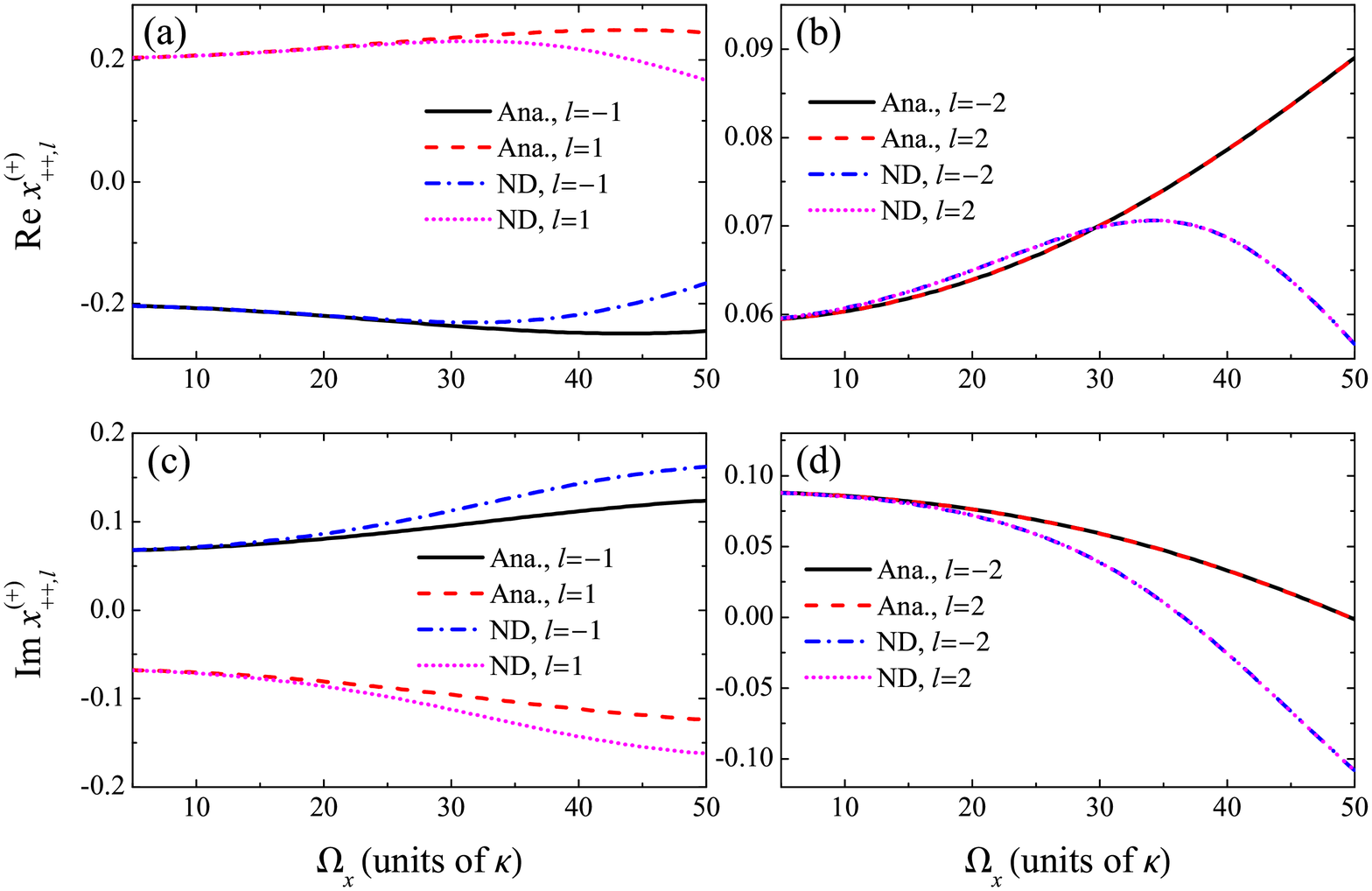}
  \caption{Transition matrix elements $x^{(+)}_{++,l}$ versus driving strength $\Omega_x$, calculated from the analytical method and the numerical method based on the ND of the Floquet Hamiltonian for $p=2$, $\delta=0$, $\Omega_z=\omega_z=40\kappa$, $\phi=\pi/2$, and $r=1$.}\label{fig5}
\end{figure}

\begin{figure}
  \includegraphics[width=1\columnwidth]{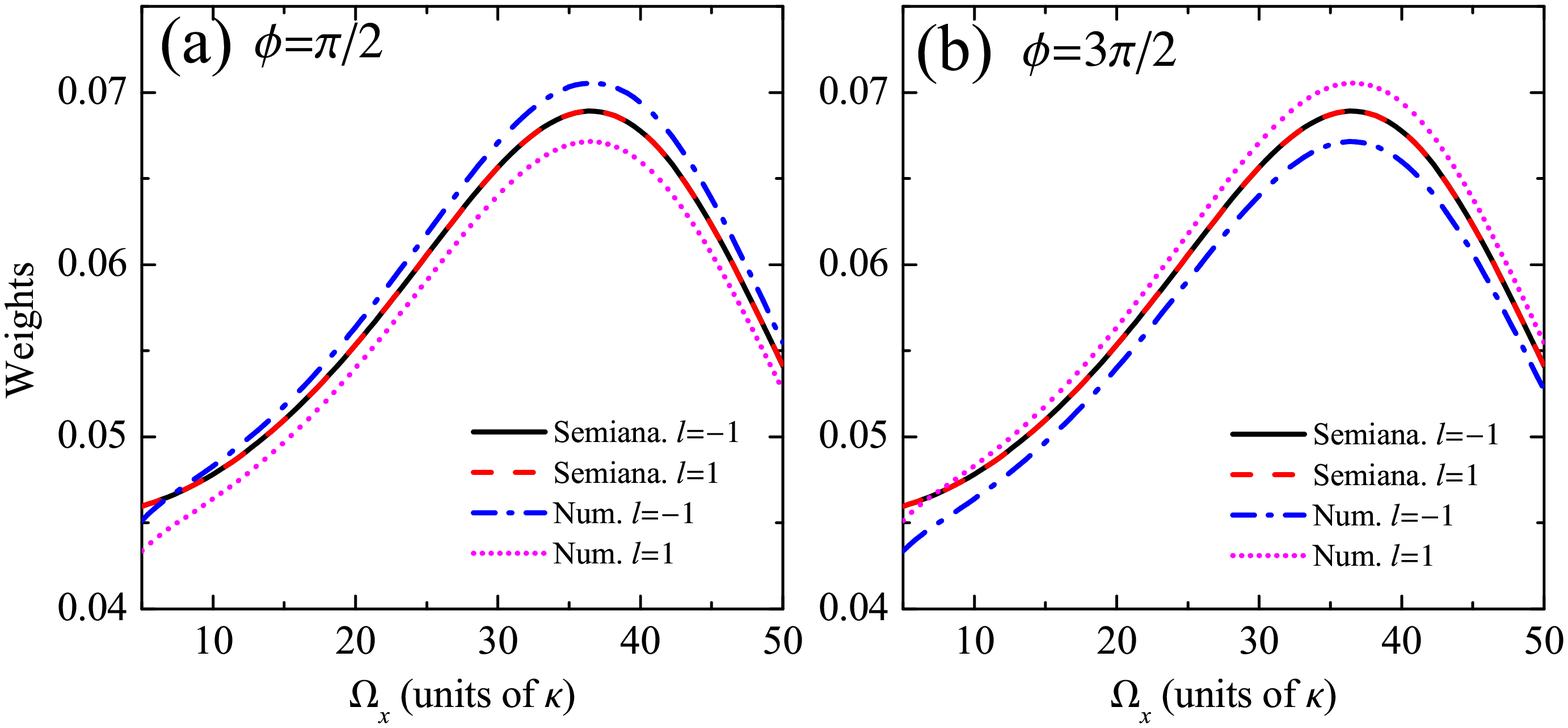}
  \caption{Weights of emission lines at $\Delta=\pm\omega_z$ versus driving strength $\Omega_x$, calculated from the semianalytical method and the FL method, for $p=2$, $\delta=0$, $\Omega_z=\omega_z=40\kappa$, $r=1$, and two values of $\phi$. ``Semiana.'' denotes the semianalytical result.}\label{fig6}
\end{figure}

Let us explore how the analytical spectrum becomes symmetric in the absence of the generalized parity of the Floquet states. To this end, we show that the relation~(\ref{eq:identity2}) can originate from the identities different from Eq.~(\ref{eq:identity}). Based on the results from the Van Vleck perturbation theory, we analytically derive the identities for the transition matrix elements in the case of vanishing detuning and even $p$. The derivation are given in Appendix~\ref{AppB}. When $p$ is even, $\delta=0$, and $\phi=\left(1/2+n\right)\pi$ ($n=0,\pm1,\pm2,\ldots$), we find that the following relations hold for arbitrary integer $l$:
\begin{eqnarray}
  x^{(+)}_{++,-l}& = &(-1)^{l}x^{(+)}_{++,l}, \label{eq:x11eq}\\
  x^{(+)}_{-+,-l}& = &-(-1)^{l}e^{-i2\theta_{0}}x^{(+)}_{+-,l}, \label{eq:x21eq}
\end{eqnarray}
where $\theta_0$ is a phase defined in Eq.~(\ref{eq:theta0}). Although the relations (\ref{eq:x11eq}) and (\ref{eq:x21eq}) are derived based on the perturbation theory, it is straightforward to show that they hold in the nonperturbative regimes. In Fig.~\ref{fig5}, we calculate $x^{(+)}_{++,l}$ $(l=\pm1,\pm2)$ with the variation of $\Omega_x$ by using the analytical and ND methods. We see that the deviation between the analytical and numerical results becomes larger and larger as $\Omega_x$ increases, which is due to the breakdown of the perturbation calculation. Nevertheless, $x^{(+)}_{++,l}$ obtained by the ND method still satisfies Eq.~(\ref{eq:x11eq}). This suggests that the relations~(\ref{eq:x11eq}) and (\ref{eq:x21eq}) are not limited to the perturbative regimes. More importantly,
it follows from the identities~(\ref{eq:x11eq}) and~(\ref{eq:x21eq}) that $|x_{\alpha\beta,l}^{(+)}|=|x_{\beta\alpha,-l}^{(+)}|$, which leads to the symmetry of the formal spectrum~(\ref{eq:sfunsa}). That is to say, without the generalized parity of the Floquet states, the relation~(\ref{eq:identity2}) can also be established from other kinds of the identities for the transition matrix elements instead of the generalized-parity-induced identity~(\ref{eq:identity}) under certain conditions.

The discrepancy in the symmetry predicted by the analytical and numerical methods shown in Fig.~\ref{fig3}(b) indicates that the relations~(\ref{eq:x11eq}) and ~(\ref{eq:x21eq}) cannot guarantee the symmetry of the spectrum without the secular approximation. To further verify this, in Fig.~\ref{fig6}, we use semianalytical and FL numerical methods to calculate the weights of the emission lines at $\Delta=\pm\omega_z$ as the increasing of $\Omega_x$ for $p=2$, $\delta=0$, and two values of $\phi$. It is evident that the weights calculated from the semianalytical method (solid and dashed lines) are the same while the weights from the numerical method (dot-dashed and dotted lines) are unequal, indicating that the semianalytical spectrum is symmetric but the numerical spectrum is not symmetric.
The present results illustrate that that provided the relation~(\ref{eq:identity2}) is established in the absence of the generalized parity, the secular approximation can induce artifact symmetry that vanishes if such approximation is not invoked.

Apart from the biharmonic modulation, we find that the conditions for the symmetry and asymmetry of the spectrum, which are derived based on the generalized parity, are applicable to the simple harmonic and multiharmonic modulation cases. For the simple harmonic modulation $f(t)=\Omega_{z}\cos(\omega_{z}t)$, $f(t)=-f(t+T/2)$ is met. Therefore, the symmetry and asymmetry of the spectrum is uniquely controlled by the detuning $\delta$, which simply interprets the detuning-dependent symmetry of the spectrum. Specifically, the spectrum is expected to be symmetric when $\delta=0$ and asymmetric when $\delta\neq0$. This is consistent with the findings of previous studies~\cite{PhysRevA.93.033812,PhysRevA.96.063812,PhysRevA.97.033817}. For the multiharmonic modulation $f(t)=\sum_{p=1}^{N} \Omega_{z,p}\cos(p\omega_z t+\phi_p)]$, where $\Omega_{z,p}$ and $\phi_p$ are the amplitude and phase of the $p$th harmonic, respectively, either $f(t)=-f(t+T/2)$ or $f(t)\neq-f(t+T/2)$ can be met, similarly to the biharmonic case. We have calculated the spectrum with the FL and semianalytical methods for the cases of $N=3$, $N=4$, and $N=5$. The results (not shown here) further confirm that the symmetry and asymmetry of spectrum fundamentally originate from the presence and absence of the generalized parity of the Floquet states, respectively.

\section{Conclusions}
In summary, we have studied the fundamental origin of the symmetry of the resonance fluorescence from the two-level system subjected to a periodic frequency modulation and a near-resonant high-frequency monochromatic excitation by using both analytical and numerical methods based on the Floquet theory. In such a driven two-level system, we have found that the generalized parity of Floquet states plays a fundamental role in the symmetry of the spectrum. Specifically, the generalized parity guarantees the symmetry of the spectrum. On the other hand, when the generalized parity is broken, the spectrum becomes asymmetric. This has been illustrated in the context of the biharmonic modulation, the parameters of which can be tuned to induce or break the generalized parity. For the biharmonic modulation, we find that when $\delta=0$ and $f(t)=-f(t+T/2)$, the generalized parity exists and the spectrum is symmetric. When $\delta+f(t)\neq-[\delta+f(t+T/2)]$, the generalized parity is broken and the spectrum is found to be asymmetric. Interestingly, we can obtain pronouncedly asymmetric spectrum by requiring the modulation $f(t)\neq-f(t+T/2)$ even though $\delta=0$. Moreover, these conditions for the symmetry and asymmetry of the spectrum are found to be applicable to the simple harmonic and multiharmonic modulation cases.
In addition, we illustrated that the secular approximation may induce artifact symmetry that vanishes if the secular approximation is avoided under certain conditions. The present study gives a deep insight into the origin
of the symmetry of the spectrum and reveals a simple relation between the symmetry of the spectrum and the generalized parity of the Floquet states.

\begin{acknowledgments}
This work was supported by the National Natural Science
Foundation of China (Grants No. 11647082, No. 11774311, No. 11774226, and No. 11874260).
\end{acknowledgments}

\appendix
\section{Derivation of symmetry of the spectrum without the secular approximation}
\label{AppC}
The master equation can be rewritten in a matrix form
\begin{equation}
\frac{d}{dt}\vec{\tilde{\rho}}(t)={\cal L}(t)\vec{\tilde{\rho}}(t).\label{eq:blocheq}
\end{equation}
Here the vector is defined as
\begin{equation}
\vec{\tilde{\rho}}(t)=(\langle\tilde{\sigma}_{+}(t)\rangle,\langle\tilde{\sigma}_{-}(t)\rangle,\langle\tilde{\pi}_{+}(t)\rangle,\langle\tilde{\pi}_{-}(t)\rangle)^{{\rm T}},
\end{equation}
where $\pi_{\pm}=(1\pm\sigma_{z})/2$ and $\langle \tilde{\hat{o}}(t)\rangle \equiv{\rm Tr}[\hat{o}\tilde{\rho}(t)]$.
The superoperator ${\cal L}(t)$ in the Liouville space spanned by the matrix bases $\{\sigma_{\pm},\pi_{\pm}\}$ is given by
\begin{equation}
{\cal L}(t)=\left(\begin{array}{cccc}
i[\delta+f(t)]-\frac{\kappa}{2} & 0 & -\frac{i\Omega_{x}}{2} & \frac{i\Omega_{x}}{2}\\
0 & -i[\delta+f(t)]-\frac{\kappa}{2} & \frac{i\Omega_{x}}{2} & \frac{-i\Omega_{x}}{2}\\
\frac{-i\Omega_{x}}{2} & \frac{i\Omega_{x}}{2} & -\kappa & 0\\
\frac{i\Omega_{x}}{2} & \frac{-i\Omega_{x}}{2} & \kappa & 0
\end{array}\right).
\end{equation}

If $\delta+f(t)=-[\delta+f(t+T/2)]$, in which case the Hamiltonian
is invariant under the generalized parity transformation, one readily
finds that
\begin{equation}
{\cal T}{\cal L}(t+T/2){\cal T}={\cal L}(t),\label{eq:Mtparity}
\end{equation}
where the transformation matrix is given by
\begin{equation}
{\cal T}=\left(\begin{array}{cccc}
0 & 1 & 0 & 0\\
1 & 0 & 0 & 0\\
0 & 0 & -1 & 0\\
0 & 0 & 0 & -1
\end{array}\right),
\end{equation}
 and ${\cal T}^{2}=I$ with $I$ being the identity matrix. Similarly
to the Hamiltonian, the matrix ${\cal L}(t)$ is invariant under
the transformation defined in Eq.~(\ref{eq:Mtparity}), which can be regarded as the generalized parity transformation in the Liouville space, similarly to that defined in Eq.~(\ref{eq:hpt}) of the main text.

Let us derive the specific property of the steady state in the long-time
limit [as $\det {\cal L}(t)=0$, there exists a nontrivial steady state].
It follows from Eq.~(\ref{eq:blocheq}) that
\begin{equation}
\frac{d}{dt}\vec{\tilde{\rho}}(t+T/2)={\cal L}(t+T/2)\vec{\tilde{\rho}}(t+T/2),
\end{equation}
which leads to
\begin{widetext}
\begin{eqnarray}
\frac{d}{dt}{\cal T}\vec{\tilde{\rho}}(t+T/2) & = & {\cal T}{\cal L}(t+T/2){\cal T}{\cal T}\vec{\tilde{\rho}}(t+T/2)  =  {\cal L}(t){\cal T}\vec{\tilde{\rho}}(t+T/2),
\end{eqnarray}
\end{widetext}
which means that ${\cal T}\vec{\tilde{\rho}}(t+T/2)=c\vec{\tilde{\rho}}(t)$, owing
to the uniqueness of solutions of the differential equation. On using
the fact that $\vec{\tilde{\rho}}(t)=\vec{\tilde{\rho}}(t+T)$
as $t\rightarrow\infty$ because of ${\cal L}(t)={\cal L}(t+T)$, we find that $c$
may be either $+1$ or $-1$. It is easy to prove by contradiction that $c=-1$. Suppose that $c=1$, yielding $\langle\tilde{\pi}_{+}(t+T/2)\rangle=-\langle\tilde{\pi}_{+}(t)\rangle$.
However, if one considers $\delta+f(t)=0$ in which case ${\cal L}(t)$ is time independent while Eq.~(\ref{eq:Mtparity})
still holds, the steady state becomes time independent
and one gets $\langle\tilde{\pi}_{+}(t)\rangle=\langle\tilde{\pi}_{+}(t+T/2)\rangle$.
By contradiction, one finds that $c=-1$. Consequently, in the steady-state
limit, we have
\begin{equation}
{\cal T}\vec{\tilde{\rho}}(t+T/2)=-\vec{\tilde{\rho}}(t)\quad(t\rightarrow\infty).\label{eq:ss}
\end{equation}

Next, let us derive the property of the principal matrix solution $\Pi(t,t^{\prime})$ of the master equation,
which solves the differential equation
\begin{equation}
\frac{d}{dt}\Pi(t,t^{\prime})={\cal L}(t)\Pi(t,t^{\prime}),
\end{equation}
with the initial condition $\Pi(t^{\prime},t^{\prime})=I$. It is
straightforward to show that
\begin{widetext}
\begin{eqnarray}
\frac{d}{dt}{\cal T}\Pi(t+T/2,t^{\prime}+T/2){\cal T} & = & {\cal T}{\cal L}(t+T/2){\cal T}{\cal T}\Pi(t+T/2,t^{\prime}+T/2){\cal T} =  {\cal L}(t){\cal T}\Pi(t+T/2,t^{\prime}+T/2){\cal T},
\end{eqnarray}
namely, ${\cal T}\Pi(t+T/2,t^{\prime}+T/2){\cal T}$ satisfies the
same differential equation and the same initial condition as $\Pi(t,t^{\prime})$.
As a result, we simply have
\begin{equation}
{\cal T}\Pi(t+T/2,t^{\prime}+T/2){\cal T}=\Pi(t,t^{\prime}).\label{eq:Piparity}
\end{equation}

According to the quantum regression theory~\cite{PhysRev.188.1969}, the two-time correlation
functions
\begin{equation}
\vec{\tilde{g}}(t,t^{\prime})=(\langle\tilde{\sigma}_{+}(t)\tilde{\sigma}_{-}(t^{\prime})\rangle,\langle\tilde{\sigma}_{-}(t)\tilde{\sigma}_{-}(t^{\prime})\rangle,\langle\tilde{\pi}_{+}(t)\tilde{\sigma}_{-}(t^{\prime})\rangle,\langle\tilde{\pi}_{-}(t)\tilde{\sigma}_{-}(t^{\prime})\rangle)^{{\rm T}}
\end{equation}
satisfy the same equation as $\vec{\tilde{\rho}}(t)$, however,
with a different initial condition
\begin{equation}
\vec{\tilde{g}}(t^{\prime},t^{\prime})=(\langle\tilde{\pi}_{+}(t^{\prime})\rangle,0,0,\langle\tilde{\sigma}_{-}(t^{\prime})\rangle)^{{\rm T}}.
\end{equation}
Similarly, another set of two-time correlation functions
\begin{equation}
\vec{\tilde{G}}(t,t^{\prime})=(\langle\tilde{\sigma}_{+}(t^{\prime})\tilde{\sigma}_{+}(t)\rangle,\langle\tilde{\sigma}_{+}(t^{\prime})\tilde{\sigma}_{-}(t)\rangle,\langle\tilde{\sigma}_{+}(t^{\prime})\tilde{\pi}_{+}(t)\rangle,\langle\tilde{\sigma}_{+}(t^{\prime})\tilde{\pi}_{-}(t)\rangle)^{{\rm T}}
\end{equation}
also satisfy the same differential equation as $\vec{\tilde{g}}(t,t^{\prime})$
but with the initial condition
\begin{equation}
\vec{\tilde{G}}(t^{\prime},t^{\prime})=(0,\langle\tilde{\pi}_{+}(t^{\prime})\rangle,0,\langle\tilde{\sigma}_{+}(t^{\prime})\rangle)^{{\rm T}}.
\end{equation}
Using Eq. (\ref{eq:ss}), we have
\begin{equation}
{\cal T}\vec{\tilde{g}}(t^{\prime},t^{\prime})=\left(\begin{array}{c}
0\\
\langle\tilde{\pi}_{+}(t^{\prime})\rangle\\
0\\
-\langle\tilde{\sigma}_{-}(t^{\prime})\rangle
\end{array}\right)=\left(\begin{array}{c}
0\\
\langle\tilde{\pi}_{+}(t^{\prime}+T/2)\rangle\\
0\\
\langle\tilde{\sigma}_{+}(t^{\prime}+T/2)\rangle
\end{array}\right)=\vec{\tilde{G}}\left(t^{\prime}+\frac{T}{2},t^{\prime}+\frac{T}{2}\right)\quad(t^{\prime}\rightarrow\infty).
\end{equation}
\end{widetext}
In the steady-state limit, the correlation functions are found to have the following relation
\begin{eqnarray}
\vec{\tilde{g}}(t,t^{\prime}) & = & \Pi(t,t^{\prime})\vec{\tilde{g}}(t^{\prime},t^{\prime})\nonumber \\
 & = & {\cal T}\Pi\left(t+\frac{T}{2},t^{\prime}+\frac{T}{2}\right){\cal T}\vec{\tilde{g}}(t^{\prime},t^{\prime})\nonumber \\
 & = & {\cal T}\Pi\left(t+\frac{T}{2},t^{\prime}+\frac{T}{2}\right)\vec{\tilde{G}}\left(t^{\prime}+\frac{T}{2},t^{\prime}+\frac{T}{2}\right)\nonumber \\
 & = & {\cal T}\vec{\tilde{G}}\left(t+\frac{T}{2},t^{\prime}+\frac{T}{2}\right)\quad(t^\prime\rightarrow\infty).
\end{eqnarray}
It follows that as $t^{\prime}\rightarrow\infty$,
\begin{eqnarray}
\langle\tilde{\sigma}_{+}(t)\tilde{\sigma}_{-}(t^{\prime})\rangle & = & \langle\tilde{\sigma}_{+}(t^{\prime}+T/2)\tilde{\sigma}_{-}(t+T/2)\rangle\nonumber \\
 & = &\langle\tilde{\sigma}_{+}(t+T/2)\tilde{\sigma}_{-}(t^{\prime}+T/2)\rangle^{\ast}.\label{eq:cfr}
\end{eqnarray}

In the steady-state limit, the first-order correlation function depends
explicitly on time $t^{\prime}$, however, the $t^{\prime}$ dependence
can be eliminated by setting $t=\tau+t^{\prime}$ and integrating
over $t^{\prime}$ (because the contributions of $t^{\prime}$-dependent
terms are negligible to a long-time observation), yielding the $\tau$-dependent first-order correlation function
\begin{eqnarray}
\bar{\tilde{g}}_{1}(\tau) & \equiv & \frac{1}{T}\int_{0}^{T}\lim_{t^{\prime}\rightarrow\infty}\langle\tilde{\sigma}_{+}(\tau+t^{\prime})\tilde{\sigma}_{-}(t^{\prime})\rangle dt^{\prime}\nonumber \\
 & = & \frac{1}{T}\int_{0}^{T}\lim_{t^{\prime}\rightarrow\infty}\langle\tilde{\sigma}_{+}(\tau+t^{\prime}+T/2)\tilde{\sigma}_{-}(t^{\prime}+T/2)\rangle^{\ast}dt^{\prime}\nonumber \\
 & = & \frac{1}{T}\int_{T/2}^{T+T/2}\lim_{t^{\prime}\rightarrow\infty}\langle\tilde{\sigma}_{+}(\tau+t^{\prime})\tilde{\sigma}_{-}(t^{\prime})\rangle^{\ast}dt^{\prime}\nonumber \\
 & = & \frac{1}{T}\int_{0}^{T}\lim_{t^{\prime}\rightarrow\infty}\langle\tilde{\sigma}_{+}(\tau+t^{\prime})\tilde{\sigma}_{-}(t^{\prime})\rangle^{\ast}dt^{\prime}\nonumber \\
 & = & \bar{\tilde{g}}_{1}^{\ast}(\tau),
\end{eqnarray}
where we used relation~(\ref{eq:cfr}) and the fact that $\langle\tilde{\sigma}_{+}(\tau+t^{\prime}+T)\tilde{\sigma}_{-}(t^{\prime}+T)\rangle^{\ast}=\langle\tilde{\sigma}_{+}(\tau+t^{\prime})\tilde{\sigma}_{-}(t^{\prime})\rangle^{\ast}$ as $t^\prime\rightarrow\infty$.
This means that the generalized parity guarantees that the correlation
function is a real-valued function of $\tau$ in the rotating frame and thus results
in the symmetry of the spectrum when $\delta+f(t)=-[\delta+f(t+T/2)]$.
This is consistent with the prediction from the spectrum~(\ref{eq:sfunsa}).

In general, it is a formidable task to show that the spectrum is asymmetric when $\delta+f(t)\neq-[\delta+f(t+T/2)]$
with or without the secular approximation. Nevertheless, from the above derivation, one readily notes that the generalized parity plays an important role in determining the symmetry of the spectrum. Consequently, if such parity breaks, it is not difficult to imagine that the symmetry of the spectrum also breaks trivially if there is no other symmetry-inducing mechanism.

\section{Analytical calculation of quasienergies and transition matrix elements in the biharmonic modulation case}\label{AppA}
We use the Van Vleck perturbation theory~\cite{cohen1992atom,PhysRevA.81.022117} to analytically calculate
the quasienergies and transition matrix elements $x_{\alpha\beta,l}^{(+)}$ for the
biharmonic modulation, which leads to the analytical fluorescence
spectrum. Since we are interested in the regime of $\Omega_{z},\,\omega_{z}\gg\Omega_{x}$, which is accessible in the experiment~\cite{li2013motional},
we use $\Omega_{x}$ as the perturbation parameter. We first transform
Eq.~(\ref{eigEQ}) with the unitary transformation
\begin{equation}
e^{S(t)}[\tilde{H}(t)-i\partial_{t}]e^{-S(t)}e^{S(t)}|\tilde{u}_{\alpha}(t)\rangle=\tilde{\varepsilon}_{\alpha}e^{S(t)}|\tilde{u}_{\alpha}(t)\rangle,
\end{equation}
where
\begin{equation}
S(t)=i\frac{\Omega_{z}}{2\omega_{z}}\left\{ \sin(\omega_{z}t)+\frac{r}{p}[\sin(p\omega_{z}t+\phi)-\sin\phi]\right\} \sigma_{z}.
\end{equation}
We can define the transformed Floquet states and transformed Hamiltonian as follows:
\begin{equation}
|u_{\alpha}^{\prime}(t)\rangle=e^{S(t)}|\tilde{u}_{\alpha}(t)\rangle,
\end{equation}
\begin{eqnarray}
H^{\prime}(t) & = & e^{S(t)}[\tilde{H}(t)-i\partial_{t}]e^{-S(t)}\nonumber \\
 & = & \frac{1}{2}\delta\sigma_{z}+\frac{1}{2}\sum_{l}(f_{l}\sigma_{+}+f_{-l}^{\ast}\sigma_{-})e^{il\omega_{z}t},
\end{eqnarray}
where
\begin{equation}
f_{l}=\Omega_{x}F_{l},
\end{equation}
and
\begin{eqnarray}
F_{l} & = & \frac{1}{T}\int_{0}^{T}e^{i\frac{\Omega_{z}}{\omega_{z}}\left\{ \sin(\omega_{z}t)+\frac{r}{p}[\sin(p\omega_{z}t+\phi)-\sin\phi]\right\} -il\omega_{z}t}dt\nonumber \\
 & = & e^{-i\Theta}\sum_{k}J_{k}\left(\frac{r\Omega_{z}}{p\omega_{z}}\right)J_{l-kp}\left(\frac{\Omega_{z}}{\omega_{z}}\right)e^{ik\phi},\label{eq:Fl}
\end{eqnarray}
with $\Theta=\frac{r\Omega_{z}}{p\omega_{z}}\sin\phi$ and $J_{k}(z)$ being the Bessel function of the first kind. To proceed, we introduce an extended Hilbert space in which the time-dependent Floquet Hamiltonian $H^\prime(t)-i\partial_t$ becomes time independent~\cite{PhysRevA.7.2203}. One readily introduces
the Fourier basis $|l\rangle\equiv\exp(il\omega_{z}t)$ and inner
product $\langle l|n\rangle\equiv\frac{1}{T}\int_{0}^{T}\exp[i(n-l)\omega_{z}t]dt=\delta_{l,n}$, where $\delta_{l,n}$ is the Kronecker delta function. Denoting $|\uparrow\rangle$ and $|\downarrow\rangle$ as the eigenstates for $\sigma_z$ with the eigenvalues $+1$ and $-1$, respectively, one gets
the composite bases $|\uparrow(\downarrow),l\rangle=|\uparrow(\downarrow)\rangle\otimes|l\rangle$. In the extended Hilbert space spanned by such bases, we can obtain the explicit form of the Floquet Hamiltonian, which is written as
\begin{eqnarray}
H_{{\cal F}}^{\prime} & = & H^{\prime}(t)-i\partial_{t}\nonumber \\
 & = & \frac{1}{2}\delta\sigma_{z}+\sum_{n}n\omega_{z}|n\rangle\langle n|+\frac{1}{2}\sum_{n,l}(f_{l}\sigma_{+}+f_{-l}^{\ast}\sigma_{-})\nonumber\\
 &   & \otimes|n+l\rangle\langle n|.
\end{eqnarray}

The Floquet Hamiltonian has an infinite size and is difficult to be diagonalized
exactly in analytical calculation. To carry out perturbation calculation, we transform the Floquet Hamiltonian with
a further unitary transformation with the Hermitian generator $K$,
leading to
\begin{eqnarray}
H_{{\cal F}}^{\prime\prime} & = & e^{iK}H_{{\cal F}}^{\prime}e^{-iK}\nonumber \\
 & = & H_{\cal F}^{\prime}+[iK,H_{{\cal F}}^{\prime}]+\frac{1}{2!}[iK,[iK,H_{{\cal F}}^{\prime}]]+\ldots,
\end{eqnarray}
where the explicit form of $K$ is to be determined by requiring $H_{{\cal F}}^{\prime\prime}$
to be block diagonal. The generator is expanded as
\begin{equation}
K=K^{(1)}+K^{(2)}+K^{(3)}+\ldots,
\end{equation}
where the superscripts indicate the orders in the perturbation. We
use $H_{0}=\frac{1}{2}\delta\sigma_{z}+\sum_{n}n\omega_{z}|n\rangle\langle n|$
and $V=\frac{1}{2}\sum_{n,l}(f_{l}\sigma_{+}+f_{-l}^{\ast}\sigma_{-})\otimes|n+l\rangle\langle n|$
as the dominate and perturbation components, respectively. Up to the
second order in $\Omega_{x}$, we have
\begin{eqnarray}
H_{{\cal F}}^{\prime\prime}&\simeq& H_{0}+V+[iK^{(1)},H_{0}]+[iK^{(1)},V]+[iK^{(2)},H_{0}]\nonumber\\
&   &+\frac{1}{2}[iK^{(1)},[iK^{(1)},H_{0}]].
\end{eqnarray}

Next, we discuss under which condition the transformed Hamiltonian may reasonably be block diagonal. For the dominate component $H_0$, we simply have  $H_0|\uparrow(\downarrow),n\rangle=[+(-)\delta/2+n\omega_z]|\uparrow(\downarrow),n\rangle\equiv\tilde{\varepsilon}^{(0)}_{+(-),n}|\uparrow(\downarrow),n\rangle$. Provided that $\tilde{\varepsilon}^{(0)}_{+,n}-\tilde{\varepsilon}^{(0)}_{-,n+m}=\delta-m\omega_{z}\approx0$, we have a subspace spanned by two almost degenerate
unperturbed states $|\uparrow,n\rangle$ and $|\downarrow,n+m\rangle$, where $n$ is an arbitrary integer and $m$ is the integer nearest to $\delta/\omega_z$. The projection onto such a subspace is realized by
the operator:
\begin{equation}
\Pi_{n}=|\uparrow,n\rangle\langle\uparrow,n|+|\downarrow,n+m\rangle\langle\downarrow,n+m|.
\end{equation}
The eigenvalues of the dominate component $H_0$ in the $n$th subspace are well-separated from those in the $(n+l)$th subspace as long as $|l\omega_z|\gg|\delta-m\omega_z|$ for any $l\neq0$. Moreover, if we assume that
\begin{equation}
  |\langle\uparrow,n|V|\downarrow,n+l+m\rangle|\ll|\tilde{\varepsilon}^{(0)}_{+,n}-\tilde{\varepsilon}^{(0)}_{-,n+l+m}|, \label{eq:convvp}
\end{equation}
which is simply $|f_{-l-m}/2|\ll|l\omega_z|$, the transitions between the states in the different subspaces can be neglected up to a certain order in the perturbation~\cite{cohen1992atom}, yielding
the following condition
\begin{equation}
\Pi_{n}H_{{\cal F}}^{\prime\prime}\Pi_{l}=0,\label{eq:Keq1}
\end{equation}
for $n\neq l$. Therefore, $H_{{\cal F}}^{\prime\prime}$ is block diagonal. The second condition that $K$ cannot have matrix
elements inside each subspace of two almost degenerate states is also
assumed, i.e.,
\begin{equation}
\Pi_{n}K\Pi_{n}=0.\label{eq:Keq2}
\end{equation}
The generator can now be fully determined via Eqs.~(\ref{eq:Keq1})
and (\ref{eq:Keq2}). The nonvanishing elements of $K^{(1)}$ and
$K^{(2)}$are given by
\begin{equation}
\langle\uparrow,n|iK^{(1)}|\downarrow,l\rangle=\frac{1}{2}\frac{f_{n-l}}{\delta+(n-l)\omega_{z}},
\end{equation}
\begin{equation}
\langle\downarrow,l|iK^{(1)}|\uparrow,n\rangle=-\frac{1}{2}\frac{f_{n-l}^{\ast}}{\delta+(n-l)\omega_{z}},
\end{equation}
for $n-l\neq-m$, and
\begin{widetext}
\begin{eqnarray}
\langle\uparrow,n|iK^{(2)}|\uparrow,l\rangle & = & \frac{1}{4(n-l)\omega_{z}}\left\{ \sum_{k\neq n+m,l+m}\frac{f_{n-k}f_{l-k}^{\ast}}{2}\left[\frac{1}{\delta+(n-k)\omega_{z}}+\frac{1}{\delta+(l-k)\omega_{z}}\right]\right.\nonumber \\
 &  & \left.+\frac{f_{l-n-m}^{\ast}f_{-m}}{\delta+(l-n-m)\omega_{z}}+\frac{f_{n-l-m}f_{-m}^{\ast}}{\delta+(n-l-m)\omega_{z}}\right\} ,
\end{eqnarray}
\begin{eqnarray}
\langle\downarrow,n|iK^{(2)}|\downarrow,l\rangle & = & -\frac{1}{4(n-l)\omega_{z}}\left\{ \sum_{k\neq l-m,n-m}\frac{f_{k-n}^{\ast}f_{k-l}}{2}\left[\frac{1}{\delta+(k-n)\omega_{z}}+\frac{1}{\delta+(k-l)\omega_{z}}\right]\right.\nonumber \\
 &  & +\left.\frac{f_{l-n-m}^{\ast}f_{-m}}{\delta+(l-n-m)\omega_{z}}+\frac{f_{n-l-m}f_{-m}^{\ast}}{\delta+(n-l-m)\omega_{z}}\right\} ,
\end{eqnarray}
for $n\neq l$. The rest elements of $K^{(1)}$ and $K^{(2)}$ are vanishing.

The transformed Hamiltonian have the $2\times2$ submatrix $H_{{\cal F}}^{\prime\prime(n)}$
in the diagonal, which reads~\cite{cohen1992atom}
\begin{eqnarray}
H_{{\cal F}}^{\prime\prime(n)} & = & H_{0}\Pi_{n}+\Pi_{n}V\Pi_{n}+\frac{1}{2}\Pi_{n}[iK^{(1)},V]\Pi_{n}\nonumber \\
 & = & \left(\begin{array}{cc}
\frac{\delta}{2}+n\omega_{z}+\sum_{j\neq-m}\frac{|f_{j}|^{2}}{4(\delta+j\omega_{z})} & \frac{f_{-m}}{2}\\
\frac{f_{-m}^{\ast}}{2} & -\frac{\delta}{2}+(n+m)\omega_{z}-\sum_{j\neq-m}\frac{|f_{j}|^{2}}{4(\delta+j\omega_{z})}
\end{array}\right).
\end{eqnarray}
One can diagonalize the submatrix $H_{{\cal F}}^{\prime\prime(n)}$ analytically.
Its eigenvalues (quasienergies) are
\begin{equation}
\tilde{\varepsilon}_{\pm,n}=\frac{1}{2}\left(m\omega_{z}\pm\Omega_{m}\right)+n\omega_{z},
\end{equation}
where
\begin{equation}
\Omega_{m}=\sqrt{\left[\delta-m\omega_{z}+\sum_{j\neq-m}\frac{|f_{j}|^{2}}{2(\delta+j\omega_{z})}\right]^{2}+|f_{-m}|^{2}}.
\end{equation}
The eigenvectors are given by
\begin{eqnarray}
|\Psi_{+,n}^{\prime\prime}\rangle & = & u|\uparrow,n\rangle+v|\downarrow,n+m\rangle,\\
|\Psi_{-,n}^{\prime\prime}\rangle & = & v|\uparrow,n\rangle-u^{\ast}|\downarrow,n+m\rangle,
\end{eqnarray}
with
\begin{eqnarray}
u & = & \frac{f_{-m}}{|f_{-m}|}\sqrt{\frac{1}{2}\left[1+\frac{1}{\Omega_{m}}\left(\delta-m\omega_{z}+\sum_{j\neq-m}\frac{|f_{j}|^{2}}{2(\delta+j\omega_{z})}\right)\right]},\label{eq:u}\\
v & = & \sqrt{\frac{1}{2}\left[1-\frac{1}{\Omega_{m}}\left(\delta-m\omega_{z}+\sum_{j\neq-m}\frac{|f_{j}|^{2}}{2(\delta+j\omega_{z})}\right)\right]}.\label{eq:v}
\end{eqnarray}

The eigenvectors for $H_{{\cal F}}^{\prime}$ can be derived as follows:
\begin{equation}
|\Psi_{\pm,n}^{\prime}\rangle = e^{-iK}|\Psi_{\pm,n}^{\prime\prime}\rangle\simeq \left(1-iK^{(1)}-iK^{(2)}+\frac{1}{2!}iK^{(1)}iK^{(1)}\right)|\Psi_{\pm,n}^{\prime\prime}\rangle.
\end{equation}
It is straightforward to derive the explicit form of the eigenvectors,
which reads
\begin{eqnarray}
|\Psi_{+,n}^{\prime}\rangle & = & \frac{1}{{\cal {\cal N}}}\left\{ uB|\uparrow,n\rangle-\sum_{j\neq0}P_{j}|\uparrow,n+j\rangle+vB|\downarrow,n+m\rangle+\sum_{j\neq0}Q_{j}|\downarrow,n+m+j\rangle\right\} ,\\
|\Psi_{-,n}^{\prime}\rangle & = & \frac{1}{{\cal {\cal N}}}\left\{ vB|\uparrow,n\rangle+\sum_{j\neq0}Q_{-j}^{\ast}|\uparrow,n+j\rangle-u^{\ast}B|\downarrow,n+m\rangle+\sum_{j\neq0}P_{-j}^{\ast}|\downarrow,n+m+j\rangle\right\} ,
\end{eqnarray}
where
\begin{equation}
B=1-\frac{1}{8}\sum_{l\neq-m}\frac{|f_{l}|^{2}}{(\delta+l\omega_{z})^{2}},
\end{equation}
\begin{eqnarray}
P_{j} & = & \frac{f_{j-m}}{2[\delta+(j-m)\omega_{z}]}\left(v+\frac{u f_{-m}^{\ast}}{2j\omega_{z}}\right)+\frac{u}{4j\omega_{z}}\sum_{k\neq-m}\frac{f_{k+j}f_{k}^{\ast}}{\delta+k\omega_{z}},\label{eq:Pj}\\
Q_{j} & = & \frac{f_{-j-m}^{\ast}}{2[\delta-(j+m)\omega_{z}]}\left(u+\frac{v f_{-m}}{2j\omega_{z}}\right)+\frac{v}{4j\omega_{z}}\sum_{k\neq-m}\frac{f_{k-j}^{\ast}f_{k}}{\delta+k\omega_{z}},\label{eq:Qj}
\end{eqnarray}
and ${\cal N}=\sqrt{B^{2}+\sum_{j\neq0}(|P_{j}|^{2}+|Q_{j}|^{2})}$
is the normalization factor. The Floquet states $|u_{\alpha,n}^{\prime}(t)\rangle$ with the quasienergy $\tilde{\varepsilon}_{\alpha,n}$
can be derived from $|\Psi_{\alpha,n}^{\prime}\rangle$ by replacing
$|n\rangle$ with $e^{in\omega_{z}t}$.

With above results at hand, we can analytically calculate the transition matrix element
\begin{eqnarray}
x_{\alpha\beta,l}^{(+)} & = & \frac{1}{T}\int_{0}^{T}\langle\tilde{u}_{\alpha}(t)|\sigma_{\pm}|\tilde{u}_{\beta}(t)\rangle e^{-il\omega_{z}t}dt=\frac{1}{T}\int_{0}^{T}\langle u_{\alpha}^{\prime}(t)|e^{S(t)}\sigma_{+}e^{-S(t)}|u_{\beta}^{\prime}(t)\rangle e^{-il\omega_{z}t}dt\nonumber \\
 & = & \sum_{n}\frac{1}{T}\int_{0}^{T}F_{n}\langle u_{\alpha}^{\prime}(t)|\sigma_{+}|u_{\beta}^{\prime}(t)\rangle e^{i(n-l)\omega_{z}t}dt=\sum_{n}F_{n+l}\langle\Psi_{\alpha,0}^{\prime}|\sigma_{+}|\Psi_{\beta,n}^{\prime}\rangle,\label{eq:xpabl}
\end{eqnarray}
and
\begin{eqnarray}
\langle\Psi_{+,0}^{\prime}|\sigma_{+}|\Psi_{+,n}^{\prime}\rangle & = & \frac{1}{{\cal N}^{2}}\left\{ u^{\ast}vB^{2}\delta_{n,-m}-\sum_{j\neq0,n+m}P_{j}^{\ast}Q_{j-n-m}+(u^{\ast}Q_{-n-m}-vP_{n+m}^{\ast})B(1-\delta_{n,-m})\right\} ,\label{eq:xppn}
\end{eqnarray}
\begin{eqnarray}
\langle\Psi_{+,0}^{\prime}|\sigma_{+}|\Psi_{-,n}^{\prime}\rangle & = & \frac{1}{{\cal N}^{2}}\left\{ -(u^{\ast})^{2}B^{2}\delta_{n,-m}-\sum_{j\neq0,n+m}P_{j}^{\ast}P_{n+m-j}^{\ast}+2u^{\ast}P_{n+m}^{\ast}B(1-\delta_{n,-m})\right\} ,\label{eq:xpnn}
\end{eqnarray}
\begin{eqnarray}
\langle\Psi_{-,0}^{\prime}|\sigma_{+}|\Psi_{+,n}^{\prime}\rangle & = & \frac{1}{{\cal N}^{2}}\left\{ v^{2}B^{2}\delta_{n,-m}+\sum_{j\neq0,n+m}Q_{-j}Q_{j-n-m}+2vQ_{-n-m}B(1-\delta_{n,-m})\right\} , \label{eq:xnpn}
\end{eqnarray}
\begin{eqnarray}
\langle\Psi_{-,0}^{\prime}|\sigma_{+}|\Psi_{-,n}^{\prime}\rangle & = & \frac{1}{{\cal N}^{2}}\left\{ -u^{\ast}vB^{2}\delta_{n,-m}+\sum_{j\neq0,n+m}P_{j}^{\ast}Q_{j-n-m}+(vP_{n+m}^{\ast}-u^{\ast}Q_{-n-m})B(1-\delta_{n,-m})\right\} ,
\end{eqnarray}
\end{widetext}where $(1-\delta_{n,-m})$ indicates that the term vanishes for
$n=-m$. Clearly, the validity of the perturbation theory is limited to the condition (\ref{eq:convvp}). For $\delta\approx0$, roughly speaking, the above results can be justified when $r\sim1$ and $\omega_z\sim\Omega_z\gg\Omega_x$.

\section{Equalities for transition matrix elements in the vanishing detuning case}\label{AppB}
For the biharmonic modulation, we show the equalities that the transition matrix elements satisfy under the vanishing detuning condition ($\delta=0$) using the above analytical results, which helps us to understand the symmetry of the spectrum in the main text. It follows from Eq.~(\ref{eq:Fl}) that
\begin{eqnarray}
F_{-l} & = & e^{-i\Theta}\sum_{k}J_{k}\left(\frac{r\Omega_{z}}{p\omega_{z}}\right)J_{-l-kp}\left(\frac{\Omega_{z}}{\omega_{z}}\right)e^{ik\phi}\nonumber \\
 & = & (-1)^{l}e^{-i\Theta}\sum_{k}J_{k}\left(\frac{r\Omega_{z}}{p\omega_{z}}\right)(-1)^{k(p+1)}\nonumber\\
 &   & \times J_{l-kp}\left(\frac{\Omega_{z}}{\omega_{z}}\right)e^{-ik\phi},
\end{eqnarray}
where we used the relation $J_{-n}(z)=(-1)^{n}J_{n}(z)$. It is evident that when $p$
is an odd number, $p+1$ is even and thus $(-1)^{k(p+1)}=1$, leading
to
\begin{equation}
F_{-l}=(-1)^{l}e^{-i2\Theta}F_{l}^{\ast}.\label{eq:Fleq1}
\end{equation}
When $p$ is an even number, $(-1)^{k(p+1)}=(-1)^{k}$ may be either $+1$ or $-1$.
Nevertheless, we can obtain a simple relation between $F_{l}$ and
$F_{-l}$ by setting
\begin{equation}
(-1)^{k}e^{-ik\phi}=e^{ik\phi},
\end{equation}
which yields that $\phi=\left(1/2+n\right)\pi$ $(n=0,\pm1,\pm2,\ldots)$.
With an even $p$ and such values of phase, we have
\begin{equation}
F_{l}=(-1)^{l}F_{-l}.\label{eq:Fleq2}
\end{equation}
We should emphasize that Eqs.~(\ref{eq:Fleq1}) and~(\ref{eq:Fleq2}) hold under different conditions. The former is available when $p$ is odd and regardless of $\phi$ while the latter is established when $p$ is even and $\phi=(1/2+n)\pi$.

Provided that $\delta=0$, we get $m=\delta/\omega_z=0$. We define the phase of $F_{0}$ via
\begin{equation}
  F_0=e^{-i\theta_0}|F_0|.\label{eq:theta0}
\end{equation}
Together with Eqs.~(\ref{eq:u}) and~(\ref{eq:v}), we simply have
\begin{equation}
  v=ue^{i\theta_0}\label{eq:uv}
\end{equation}
with the aid of Eq.~(\ref{eq:Fleq1}) or~(\ref{eq:Fleq2}). Such an equality between $u$ and $v$ is valid only for $\delta=0$ and in the valid regime of Eq.~(\ref{eq:Fleq1}) or (\ref{eq:Fleq2}).

\begin{widetext}
\subsection{Odd $p$}
We consider that $p$ is an odd number. It follows from Eq.~(\ref{eq:Fl}) that $\theta_0=\Theta$. Using $\delta=0$ and Eqs.~(\ref{eq:Fleq1}) and~(\ref{eq:uv}), one readily gets from Eqs.~(\ref{eq:Pj}) and~(\ref{eq:Qj}) that
\begin{eqnarray}
Q_{j} & = & -\frac{f_{-j}^{\ast}}{2j\omega_{z}}\left(u+\frac{vf_{0}}{2j\omega_{z}}\right)+\frac{v}{4j\omega_{z}}\sum_{k\neq0}\frac{f_{k-j}^{\ast}f_{k}}{k\omega_{z}}\nonumber \\
 & = & \frac{(-1)^{j+1}e^{i2\Theta}f_{j}}{2j\omega_{z}}\left(u+\frac{vf_{0}^{\ast}e^{-i2\Theta}}{2j\omega_{z}}\right)+\frac{v}{4j\omega_{z}}\sum_{k\neq0}\frac{f_{-k-j}^{\ast}f_{-k}}{-k\omega_{z}}\nonumber \\
 & = & \frac{(-1)^{j+1}e^{i\Theta}f_{j}}{2j\omega_{z}}\left(v+\frac{uf_{0}^{\ast}}{2j\omega_{z}}\right)+\frac{e^{i\Theta}u}{4j\omega_{z}}\sum_{k\neq0}\frac{(-1)^{j+1}f_{k+j}f_{k}^{\ast}}{k\omega_{z}}\nonumber \\
 & = & (-1)^{j+1}e^{i\Theta}P_{j}.\label{eq:QPeq1}
\end{eqnarray}
From this relation and Eqs.~(\ref{eq:xpabl})-(\ref{eq:xnpn}), it is straightforward to show that
\begin{eqnarray}
\left[x_{-+,-l}^{(+)}\right]^{\ast} & = & \sum_{n}\frac{F_{n-l}^{\ast}}{{\cal N}^{2}}\left\{ v^{2}B^{2}\delta_{n,0}+\sum_{n\neq0,n}Q_{-j}^{\ast}Q_{j-n}^{\ast}+2vBQ_{-n}^{\ast}(1-\delta_{n,0})\right\} \nonumber \\
 & = & \sum_{n}\frac{F_{-n-l}^{\ast}}{{\cal N}^{2}}\left\{ v^{2}B^{2}\delta_{n,0}+\sum_{j\neq0,-n}Q_{-j}^{\ast}Q_{j+n}^{\ast}+2vBQ_{n}^{\ast}(1-\delta_{n,0})\right\} \nonumber \\
 & = & \sum_{n}\frac{(-1)^{n+l}F_{n+l}e^{i2\Theta}}{{\cal N}^{2}}\left\{ v^{2}B^{2}\delta_{n,0}+\sum_{j\neq0,n}Q_{j}^{\ast}Q_{n-j}^{\ast}+2vBQ_{n}^{\ast}(1-\delta_{n,0})\right\} \nonumber \\
 & = & \sum_{n}\frac{(-1)^{n+l}F_{n+l}e^{i2\Theta}}{{\cal N}^{2}}\left\{ v^{2}B^{2}\delta_{n,0}+\sum_{j\neq0,n}(-1)^{n}e^{-i2\Theta}P_{j}^{\ast}P_{n-j}^{\ast}+2vB(-1)^{n+1}e^{-i\Theta}P_{n}^{\ast}(1-\delta_{n,0})\right\} \nonumber \\
 & = & (-1)^{l}\sum_{n}\frac{F_{n+l}}{{\cal N}^{2}}\left\{ (u^{\ast})^{2}B^{2}\delta_{n,0}+\sum_{j\neq0,n}P_{j}^{\ast}P_{n-j}^{\ast}-2u^{\ast}BP_{n}^{\ast}(1-\delta_{n,0})\right\} \nonumber \\
 & = & -(-1)^{l}x_{+-,l}^{(+)}.
\end{eqnarray}
Similarly, we find that $\left[x^{(+)}_{++,-l}\right]^\ast=(-1)^{l}x^{(+)}_{++,l}$. Not surprisingly, due to the generalized parity of the Floquet states, the transition matrix elements satisfy Eq.~(\ref{eq:identity}) as long as $\delta+f(t)=-[\delta+f(t+T/2)]$. For the biharmonic modulation, such equalities are established when $p$ is odd and $\delta=0$.

\subsection{Even $p$}
We move to consider that $p$ is an even number. In such a case, the generalized parity of the Floquet states is broken even if $\delta=0$. Thus, we cannot expect that the transition matrix elements satisfy Eq.~(\ref{eq:identity}). However, we have another type of equality. With Eqs.~(\ref{eq:Fleq2}) and (\ref{eq:uv}), one gets
\begin{eqnarray}
Q_{j} & = & \frac{f_{-j}^{\ast}}{-2j\omega_{z}}\left(u+\frac{vf_{0}}{2j\omega_{z}}\right)+\frac{v}{4j\omega_{z}}\sum_{k\neq0}\frac{f_{k-j}^{\ast}f_{k}}{k\omega_{z}}\nonumber \\
 & = & \frac{(-1)^{j+1}f_{j}^{\ast}}{2j\omega_{z}}\left(u+\frac{vf_{0}}{2j\omega_{z}}\right)+\frac{v}{4j\omega_{z}}\sum_{k\neq0}\frac{(-1)^{j+1}f_{j-k}^{\ast}f_{-k}}{-k\omega_{z}}\nonumber \\
 & = & \frac{(-1)^{j+1}e^{-i\theta_{0}}f_{j}^{\ast}}{2j\omega_{z}}\left(v+\frac{u^{\ast}f_{0}}{2j\omega_{z}}\right)+\frac{e^{-i\theta_0}u^{\ast}}{4j\omega_{z}}\sum_{k\neq0}\frac{(-1)^{j+1}f_{j+k}^{\ast}f_{k}}{k\omega_{z}}\nonumber \\
 & = & (-1)^{j+1}e^{-i\theta_{0}}P_{j}^{\ast}.\label{eq:QPeq2}
\end{eqnarray}
\end{widetext}
It is straightforward to derive Eqs.~(\ref{eq:x11eq}) and~(\ref{eq:x21eq}) in the main text via Eqs.~(\ref{eq:xpabl})-(\ref{eq:xnpn}) and~(\ref{eq:QPeq2}). We stress that the conditions for establishing such relations require that $p$ is even, $\phi=(1/2+n)\pi$, and $\delta=0$.

\bibliography{parityfluo}
\end{document}